\documentclass{article}
\usepackage{authblk}
\usepackage{amsmath}
\newcommand{\eps}{\varepsilon}
\newtheorem{thm}{Theorem}
\title{Lower Bounds for Parallel Quantum Counting}
\author[1]{Paul Burchard}
\affil[1]{Goldman, Sachs \& Co.}
\begin{document}
\maketitle
\begin{abstract}
We prove a generalization of the parallel adversary method to multi-valued functions, and apply it to prove that there is no parallel quantum advantage for approximate counting.
\end{abstract}

\section{Introduction}

It is known that the Brassard-H{\o}yer-Tapp~\cite{BHT98} algorithm for approximate quantum counting is asymptotically optimal~\cite{Nay99}.
This was proved using the polynomial method~\cite{Beals01}.
However, the polynomial method is not immediately amenable to the parallel setting, where no lower bound has been published.

The optimality of Brassard-H{\o}yer-Tapp is also known to be a consequence of the adversary method of Ambainis~\cite{Amb02,Spa06,HoySpa05}.
The adversary method has been extended to the parallel setting in~\cite{Jef13}.
To handle approximate algorithms, we will need a slight generalization of the adversary method to multi-valued functions.
Finally, we prove as a corollary that there is no quantum advantage to parallelizing the approximate counting algorithm.

\section{Generalized Adversary Method}

For approximate quantum algorithms, we are no longer computing a single-valued function.
Instead, the goal is that, with high probability, the result of the measurement is an element of a set of desired values $F(x)$.
We will show that all the same arguments for the adversary method go through,
under the condition that the adversary matrix $\Gamma$ has the property that $\Gamma_{x y} = 0$ whenever $F(x) \cap F(y) \ne \emptyset$.

For the queries, we assume an oracle which can be called on $n$ qubits, and we will assume that we can call $p$ such oracles in parallel, with the desire to see how performance improves as a function of $p$.
The oracle $O_x$ can be summarized by a Boolean vector $x$ whose component $x_i$ indicates whether a result qubit will be flipped for input $i$.
Here the inputs $i$ are values up to $N=2^n$ representing computational basis states with $n$ qubits.

In building a quantum algorithm that makes parallel queries to an oracle, we can without loss of generality assume that the $p$ parallel queries are performed simultaneously, and no other computation is performed simultaneously with the oracles.
These reduction arguments are the same as in Zalka's argument~\cite{Zal99} on lower bounds for parallel search algorithms.
Thus, including the oracle result qubits but omitting ancillary qubits, the parallel oracle call is
\[
O_x^{\otimes p} |i_1,\ldots,i_p;b_1,\ldots,b_p\rangle = |i_1,\ldots,i_p; b_1\oplus x_{i_1},\ldots,b_p \oplus x_{i_p}\rangle .
\]
The algorithm then alternates such parallel oracle calls with arbitrary unitary operations:
\[
|\psi^t_x\rangle = U_t O_x^{\otimes p} U_{t-1} O_x^{\otimes p} \cdots U_1 O_x^{\otimes p} U_0 |0\rangle .
\]
If after $T$ steps, we have with high probability computed a state spanned by the basis $F(x)$, then given another oracle $y$ such that $F(x) \cap F(y) = \emptyset$, we must have that $|\psi_x^T\rangle$ and $|\psi_y^T\rangle$ are distinguishable with high probability, so that $|\langle \psi_x^T | \psi_y^T\rangle|$ is sufficiently small.

The heart of the adversary method is to monitor the progress of the algorithm in separating the results of two oracles $x$ and $y$ by watching the decay of the weighted sum of inner products of the resulting states:
\[
W^t = \sum_{x,y} \Gamma_{x y} \delta_x \delta_y \langle\psi_x^t | \psi_y^t\rangle ,
\]
where $\Gamma$ is a chosen adversary matrix for $F$, and $\delta$ is a normalized principal eigenvector of $\Gamma$.
Note that $W^0 = \lambda(\Gamma)$ is just the spectral norm of $\Gamma$.
The requirement that $\Gamma_{x y} = 0$ whenever $F(x) \cap F(y) \ne \emptyset$ means that
if the algorithm is successful, $W^T$ is a small multiple of $W^0$.

To achieve a lower bound on the algorithm, the adversary method then bounds $|W^t - W^{t+1}|$ from above.
Following the argument of~\cite{Jef13}, we consider the parallel oracle as a single serial oracle that outputs $(x_{i_1},\ldots,x_{i_p})$ on input $(i_1,\ldots,i_p)$.
This allows us to apply the original argument for the serial case to bound $|W^t - W^{t+1}|$ from above in terms of the maximum of the spectral norms of matrices corresponding to each input $(i_1,\ldots,i_p)$:
\[
\Gamma^{i_1\cdots i_p}_{x y} = \left\{
\begin{array}{ll}
\Gamma_{x y} & \text{if $(x_{i_1},\ldots,x_{i_p})\ne(y_{i_1},\ldots,y_{i_p})$} \\
0 & \text{otherwise.}
\end{array}
\right.
\]
Thus we get:

\begin{thm}
\label{thm.bound}
For any adversary matrix $\Gamma$ for a multi-valued function $F$, in a setting with $p$ parallel oracles, we have a lower bound
\[
Q_2^{p\parallel}(F) = \Omega\left(\frac{\lambda(\Gamma)}{\max_{i_1,\ldots,i_p} \lambda(\Gamma^{i_1\cdots i_p})}\right) .
\]
where
\[
\Gamma^{i_1\cdots i_p}_{x y} = \left\{
\begin{array}{ll}
\Gamma_{x y} & \text{there exists $j$ such that $x_{i_j} \ne y_{i_j}$} \\
0 & \text{otherwise.}
\end{array}
\right.
\]
\end{thm}

As in~\cite{HoySpa05}, we can reduce this to a combinatorial computation using the bound on the spectral norm for entrywise product matrices given by~\cite{Mat90}:

\begin{thm}
\label{thm.combo}
Let $X$ and $Y$ be sets of inputs to a multi-valued function $F$ such that $F(x) \cap F(y) = \emptyset$ whenever $x\in X$ and $y\in Y$,
and let $R\subseteq X \times Y$ be a relation.
Set $R^{i_1\cdots i_p} = \{ (x,y)\in R : \text{there exists $j$ such that $x_{i_j} \ne y_{i_j}$} \}$.
Let $h,h'$ denote the minimal number of ones in any row and any column of $R$, respectively,
and let $\ell,\ell'$ denote the maximal number of ones in any row and any column in any of the relations $R^{i_1\cdots i_p}$, respectively.  Then
\[
Q_2^{p\parallel}(F) = \Omega\left(\sqrt{\frac{h h'}{\ell \ell'}}\right) .
\]
\end{thm}

\section{Applications}

We can apply Theorem~\ref{thm.combo} to the problem of approximate counting.
In this case, $F(x)$ is a subset of the set of numbers equal to $|x|_1$ up to relative error $\eps/2$.

Then we can construct two sets of inputs which will never have intersecting values as
$X = \{x: |x|_1 = K\}$ and $Y = \{y: |y|_1 = (1+\eps)K\}$.
These sets have the relation $R = \{(x,y): x\in X, y\in Y, x\le y\}$.

Then we can immediately compute $h = \binom{N-K}{\eps K}$ different $y$ containing $x$
and $h' = \binom{(1+\eps)K}{K}$ different $x$ contained in $y$.
Looking at the relations $R^{i_1\cdots i_p}$, the worst case for $\ell$ is when exactly one of the indices $i_k$ is not in $x$ and the rest are in $x$, leading to exactly one $j=k$ where $x_{i_j} \ne y_{i_j}$; this gives $\ell = \binom{N-K-1}{\eps K - 1}$ different $y$.
For $\ell'$, the worst case is when all of the $i_j$ are distinct and in $y$; to upper bound the number of choices of $x$ there are $p$ ways to choose the at least one $i_k$ that is not in $x$ (so double counting when there are more than one $i_j$ not in $x$),
and $\binom{(1+\eps)K - 1}{K}$ ways to choose $x$ given that $i_k$ is not in $x$.
Thus $\ell' \le p \binom{(1+\eps)K - 1}{K}$.  As a result we get
\[
Q_2^{p\parallel}(F) = \Omega\left(\sqrt{\frac{\binom{N-K}{\eps K}\cdot\binom{(1+\eps)K}{K}}{\binom{N-K-1}{\eps K - 1}\cdot p \binom{(1+\eps)K-1}{K}}}\right) ,
\]
which simplifies to the following:

\begin{thm}
The query complexity of $p$-parallel approximate counting is
\[
Q_2^{p\parallel}(F) = \Omega\left(\frac{1}{\eps} \sqrt{\frac{N}{p K}}\right),
\]
which is tight since running $p$ disjoint parallel counters on parts of the problem achieves the bound.
\end{thm}

\section{Discussion}

These lower bounds have important practical consequences.
Approximate quantum counting and related algorithms are the basis for a quadratic speedup of Monte Carlo simulations~\cite{Mon15}, which have broad applicability across technology and finance.

In the NISQ era~\cite{Pres18}, without error correction, decoherence severely limits the quantum circuit depth, which in known approximate quantum counting algorithms is required to be inversely proportional to the desired accuracy.
One might hope that with extra qubits one could parallelize the algorithm and lower the required circuit depth.
The present result, however, shows that such parallelization only yields a classical benefit.
Instead, circuit depth must be lowered by other techniques specific to Monte Carlo simulation, such as importance sampling.

\section{Acknowledgments}

Thank you to Scott Aaronson and Ronald de Wolf for useful feedback.

\bibliography{count}

\begin{thebibliography}{10}

\bibitem{Amb02}
Andris Ambainis.
\newblock Quantum lower bounds by quantum arguments.
\newblock {\em J. Comput. Syst. Sci.}, 64(4):750--767, June 2002.

\bibitem{Beals01}
Robert Beals, Harry Buhrman, Richard Cleve, Michele Mosca, and Ronald de~Wolf.
\newblock Quantum lower bounds by polynomials.
\newblock {\em J. ACM}, 48(4):778--797, July 2001.

\bibitem{BHT98}
Gilles Brassard, Peter H{\o}yer, and Alain Tapp.
\newblock Quantum counting.
\newblock {\em Automata Languages and Programming}, 1443, 06 1998.

\bibitem{HoySpa05}
P.~H{\o}yer and R.~\v{S}palek.
\newblock Lower bounds on quantum query complexity.
\newblock {\em Bulletin of the EATCS}, 87, 2005.

\bibitem{Jef13}
Stacey Jeffery, Fr{\'e}d{\'e}ric Magniez, and Ronald de~Wolf.
\newblock Optimal parallel quantum query algorithms.
\newblock {\em Algorithmica}, 79(2):509--529, 2017.

\bibitem{Mat90}
Roy Mathias.
\newblock The spectral norm of a nonnegative matrix.
\newblock {\em Linear Algebra and its Applications}, 139:269 -- 284, 1990.

\bibitem{Mon15}
Ashley Montanaro.
\newblock Quantum speedup of monte carlo methods.
\newblock {\em Proc. Roy. Soc. Ser. A}, 471(2181), 2015.

\bibitem{Nay99}
Ashwin Nayak and Felix Wu.
\newblock The quantum query complexity of approximating the median and related
  statistics.
\newblock In {\em STOC 1999}, pages 384--393, 1999.

\bibitem{Pres18}
John Preskill.
\newblock Quantum {C}omputing in the {NISQ} era and beyond.
\newblock {\em {Quantum}}, 2:79, August 2018.

\bibitem{Spa06}
Robert \v{S}palek and Mario Szegedy.
\newblock All quantum adversary methods are equivalent.
\newblock {\em Theory of Computing}, 2:1--18, 2006.

\bibitem{Zal99}
Christof Zalka.
\newblock Grover's quantum searching algorithm is optimal.
\newblock {\em Phys. Rev. A}, 60:2746--2751, Oct 1999.

\end{thebibliography}
\bibliographystyle{plain}
\end{document}